\documentclass[twocolumn,preprintnumbers,showpacs,prr,floatfix,amsmath,amssymb,notitlepage,superscriptaddress]{revtex4-1}
\usepackage{epsfig}
\usepackage{subfigure}
\usepackage{amsmath}
\usepackage{color}
\usepackage{amssymb}
\usepackage{setspace}
\usepackage{graphicx}
\usepackage{dcolumn}
\usepackage{bm}
\usepackage{times}
\usepackage{enumerate}
\usepackage{footnote}

\input epsf

\graphicspath{{figures/}}

\begin{document}

\author{Juan G. Restrepo}
\email{juanga@colorado.edu}
\thanks{The authors contributed equally to this work.}
\affiliation{Department of Applied Mathematics, University of Colorado at Boulder, Colorado 80309, USA}

\author{Per Sebastian Skardal}
\email{persebastian.skardal@trincoll.edu} 
\thanks{The authors contributed equally to this work.}
\affiliation{Department of Mathematics, Trinity College, Hartford, CT 06106, USA}

\title{Competitive Suppression of Synchronization and Non-Monotonic Transitions in Oscillator Communities with Distributed Time Delay}


\begin{abstract}
Community structure and interaction delays are common features of ensembles of network coupled oscillators, but their combined effect on the emergence of synchronization has not been studied in detail. We study the transitions between macroscopic states in coupled oscillator systems with community structure and time delays. We show that the combination of these two properties gives rise to non-monotonic transitions, whereby increasing the global coupling strength can both inhibit and promote synchronization, yielding both desynchronization and synchronization transitions. For relatively wide parameter choices we also observe asymmetric suppression of synchronization, where communities compete to suppress one another's synchronization properties until one or more win, totally suppressing the others to effective incoherence. Using the ansatz of Ott and Antonsen we provide analytical descriptions for these transitions that confirm numerical simulations.
\end{abstract}


\maketitle

\section{Introduction}

Understanding the emergence of collective behavior in ensembles of interacting dynamical systems remains an important area of research in the nonlinear dynamics community due to synchronization's central role in a wide range of phenomena~\cite{Strogatz2003,Pikovsky2003}. Examples from both natural and engineered systems include cardiac pacemaker dynamics~\cite{Glass1988}, power grids~\cite{Skardal2015SciAdv}, and Josephson junctions~\cite{Wiesenfeld1998PRE}. The Kuramoto phase oscillator model and its variants have proven particularly useful in building an understanding of collective behavior~\cite{Kuramoto1984}, and a large body of literature has identified features that give rise to rich nonlinear behavior, including external forcing~\cite{Childs2008Chaos}, multimodal frequency distributions~\cite{Martens2009PRE}, and mixed sign coupling~\cite{Hong2011PRL}. 


In many applications, oscillators are organized into well defined communities, where oscillators in the same community are more strongly coupled with each other than with oscillators in other communities. Examples include engineered communities of synchronizing bacteria (so-called {\it biopixels}) \cite{Prindle2012Nature,Scott2016ACS}, interconnected regional or national power grid networks \cite{Nishikawa2015NJP,Kim2015NJP}, microgrids \cite{Lai2016IEEE}, and synchronization of neuronal oscillations from different brain regions \cite{Deco2011Frontiers,Sporns2016ARP}. In addition to the topological effects that community structure has on synchronization dynamics, time delays in the transmission of information from one community to another inevitably exist when network structure is related to an underlying geometry, as in the examples listed above. While the effects of community structure and time delay on the synchronization dynamics have been studied separately in the context of the Kuramoto model~\cite{Lee2009PRL,Skardal2012PRE,Yeung1999PRL,Skardal2018IJBC,Montbrio2004PRE,Barreto2008PRE,Pietras2016PRE}, their combined effect on the collective dynamics remains, with a few exceptions (e.g., Ref.~\cite{Wang2012Comm}), relatively unexplored. Here we address this gap and uncover a number of novel nonlinear behaviors.

\begin{figure}[t]
\centering
\epsfig{file =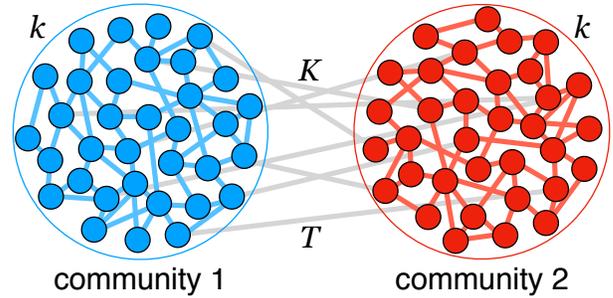, clip =,width=1\linewidth }
\caption{{\it Two oscillator communities.} Schematic illustration of two oscillator communities with intra- and inter-community coupling strengths $k$ and $K$. Interactions between oscillators in different communities have a characteristic delay time $T$.} \label{fig:00}
\end{figure}

The most notable phenomenon that arises from the combination of time delays and community structure is a number of non-monotonic synchronization transitions, where increasing the global coupling strength can either inhibit or promote synchronization. In the case of two communities (illustrated in Fig.~\ref{fig:00}) this manifests first in a desynchronization transition where locally synchronized states give way to incoherence, followed by a (subcritical) synchronization transition where incoherence gives way to global synchronization. For more than two communities a third transition occurs with incoherence giving way to local synchronization, which is then followed by the transitions described above. In addition to these non-monotonic transitions, when system the communities' parameters are chosen to be asymmetric we find that the oscillator communities compete and asymmetrically suppress one another until one or more ``win out'', totally suppressing the others' synchronization properties.  The role played by each community, i.e., which wins out and is able to remain synchronized longer, depends nonlinearly on the time delay. 

\section{Oscillator communities with distributed time delay}

We consider here a system of $C\ge2$ communities of coupled phase oscillators governed by
\begin{align}
\dot{\theta}_i^\sigma=\omega_i^\sigma + \sum_{\sigma'=1}^C\frac{K^{\sigma\sigma'}}{N_{\sigma'}}\sum_{j=1}^{N_{\sigma'}}\sin[\theta_j^{\sigma'}(t-\tau_{ij}^{\sigma\sigma'})-\theta_i^\sigma(t)],\label{eq:01}
\end{align}
where $\theta_i^\sigma$ represents the phase of oscillator $i$ in community $\sigma$, $\omega_i^\sigma$ is its natural frequency, which is assumed to be drawn from the distribution $g_\sigma(\omega)$, $K^{\sigma\sigma'}$ is the coupling strength between oscillators in communities $\sigma$ and $\sigma'$, $\tau_{ij}^{\sigma\sigma}$ is the time delay between oscillators $i$ and $j$ in communities $\sigma$ and $\sigma'$, which is assumed to be drawn from the distribution $h_{\sigma\sigma'}(\tau)$, and $N_\sigma$ is the number of oscillators in community $\sigma$. 

To begin our analysis of the dynamics of Eq.~(\ref{eq:01}) we make a few simplifying parameter choices. First we allow for two coupling strengths: $k=K^{\sigma\sigma}$ and $K=CK^{\sigma\sigma'}/2$ (for $\sigma'\ne\sigma$) denoting intra- and inter-community coupling. Next, we assume that within each community time delays are zero, i.e., $h_{\sigma\sigma}(\tau)=\delta(\tau)$, but between different communities the distribution $h_{\sigma\sigma'}$ is exponential with mean $T_{\sigma'}$, namely, $h_{\sigma\sigma'}(\tau)\propto e^{-\tau/T_{\sigma'}}$. We also consider the case where all communities are of the same size, i.e., $N_\sigma=N$ for all $\sigma$ and we assume that frequency distributions are Lorentzian with comunity-specific mean $\Omega_\sigma$ and width $\Delta$, i.e., $g_\sigma(\omega)=\Delta/\{\pi[\Delta^2+(\omega-\Omega_\sigma)^2]\}$. Next, seeking a description for the local order parameters $z_\sigma=N^{-1}\sum_{j=1}^Ne^{i\theta_j^\sigma}$ describing the degree of synchronization within each community, we apply the dimensionality reduction technique of Ott and Antonsen~\cite{Ott2008Chaos,Ott2009Chaos} to obtain the following system of reduced equations
\begin{align}
\dot{z}_\sigma&=-z_{\sigma}+i\Omega_\sigma z_\sigma + \frac{k}{2}(z_\sigma-z_\sigma^*z_\sigma^2)\label{eq:02}\\
+&\frac{K}{C}\sum_{\sigma'\ne\sigma}(w_{\sigma'}-w_{\sigma'}^*z_\sigma^2),\hskip4ex
T_\sigma \dot{w}_\sigma=z_\sigma-w_\sigma,\label{eq:03}
\end{align}
where $w_\sigma$ represents a time-delayed order parameter for community $\sigma$, with $\sigma = 1,2,\dots,C$ and the width parameter $\Delta$ of the natural frequency distributions have been scaled out. (Details of the dimensionality reduction are provided in Appendix~\ref{derivation}).

\begin{figure}[t]
\centering
\epsfig{file =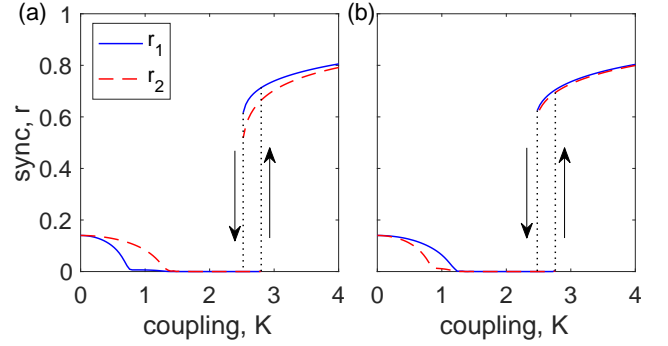, clip =,width=1\linewidth }
\caption{{\it Synchronization branches: Two communities.} Local order parameters $r_1$ (solid blue) and $r_2$ (dashed red) versus $K$ for $\kappa=0.02$, $\Omega_1=-1$, $\Omega_2=2$, and time delays  $T=1$ (a) and $0.1$ (b). Vertical dotted lines with arrows indicate hysteresis.} \label{fig:01}
\end{figure}

\subsection{Two communities} 

To illustrate the rich dynamics introduced by the interplay between time delay and hierarchical community structure in the simplest setting, we consider the case of two communities with $T_1 = T_2 = T$, illustrated in Fig.~\ref{fig:00}. We also define $\kappa = (k-2)/2$, noting that for $K=0$ each community undergoes a transition to synchronization at $k=2$. Therefore $\kappa > 0$ ($\kappa<0$) indicates that the isolated communities would be synchronized (incoherent). Equations~(\ref{eq:02})-(\ref{eq:03}) become
\begin{align}
\dot z_1 &= \kappa z_1 + i\Omega_1 z_1- (1+\kappa) z_1^* z_1^2 + \frac{K}{2}(w_2 - w_2^* z_1^2),\label{eq:04}\\
\dot z_2 &= \kappa z_2 + i\Omega_2 z_2 - (1+\kappa) z_2^* z_2^2 + \frac{K}{2}(w_1 - w_1^* z_2^2),\label{eq:05}\\
&~~~~~~~~~T \dot w_1 = z_1 - w_1, \hspace{0.5cm}T \dot w_2 = z_2 - w_2.\label{eq:06}
\end{align}
Equations~(\ref{eq:04})--(\ref{eq:06}) display some remarkable dynamics, which we illustrate in Fig.~\ref{fig:01} for parameters $\kappa = 0.02$, $\Omega_1=-1$, $\Omega_2=2$, and $T=1$, $T=0.1$ for panels (a) and (b), respectively. We plot the time-averaged values of $r_1 = |z_1|$ (solid blue) and $r_2 = |z_2|$ (dashed red) obtained from first slowly increasing $K$ from $0$ to $4$, then slowly decreasing it back to $0$. (For some parameters, $r_1$ and $r_2$ have oscillatory or chaotic dynamics, which are not the focus of this paper.) For sufficiently small $K$ both communities are partially synchronized with $r_1,r_2>0$ but are not synchronized with one another as the angles $\psi_1,\psi_2$ do not phase-lock (not shown). We call this state {\it local synchronization}. As $K$ is increased we observe that one of the communities nearly desynchronizes ($r_1\approx 0$ or $r_2\approx 0$) while the other remains synchronized, a state we call {\it asymmetric suppression}. (Although complete incoherence is not reached because of the pulling effect of the synchronized community, the transition is easy to see.) More specifically, for $T=1$ community $1$ undergoes this transition, whereas for $T=0.1$ it is community 2 that undergoes this transition. Since the oscillator communities differ only in their mean frequencies $\Omega_\sigma$ and have the same spread $\Delta=1$ (which typically dictates local synchronization properties) the identity of the desynchronizing community depends on the characteristic time delay rather than any community-specific properties. In particular, while both communities suppress one another's synchronization properties, one eventually wins out, remaining synchronized for larger values of $K$. Further increasing $K$ yields complete incoherence, i.e., both $r_1,r_2=0$, followed by a subcritical (sometimes called ``explosive''~\cite{Gardenes2011PRL}) transition to global synchronization. When $K$ is decreased the system undergoes a similarly explosive desynchronization transition to incoherence at a different coupling strength (highlighting the existence of a hysteresis loop), followed by a return to local synchronization through the asymmetric suppression state. We note that the results of the reduced equations presented here are in excellent agreement with those obtained from direct simulations of a microscopic system based on Eq.~(\ref{eq:01}) using communities of size $N=2\times10^4$, which are presented in Appendix~\ref{validation}, and similarly for results presented below.

\begin{figure}[t]
\centering
\epsfig{file =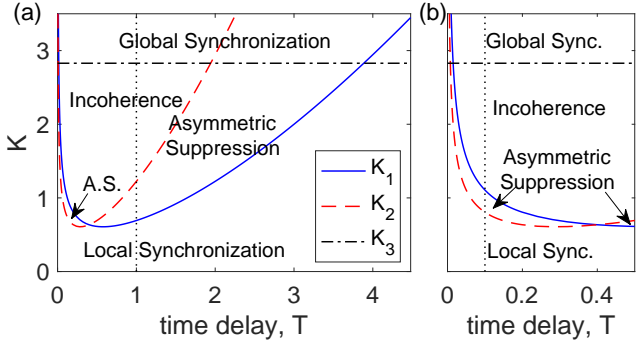, clip =,width=1\linewidth }
\caption{{\it Stability diagram: Two communities.} (a) Critical coupling values $K_1$, $K_2$, and $K_3$ (solid blue, dashed red, and dot-dashed black, respectively) given by Eqs.~(\ref{eq:09})--(\ref{eq:11}) versus time delay $T$ for $\kappa=0.02$, $\Omega_1=-1$, and $\Omega_2=2$. Stability regimes of local synchronization, asymmetric suppression, incoherence, and global synchronization are indicated between each curve. (b) Zoomed-in view of the small-$T$ region.} \label{fig:02}
\end{figure}

To better understand this sequence of bifurcations, we perform a linear stability analysis of the incoherent state $z_1 = z_2 = w_1 = w_2 = 0$. Details can be found in Appendix~\ref{linear stability analysis}. Linearizing Eqs.~(\ref{eq:04})--(\ref{eq:06}) and looking for values of $K$ where solutions have a purely imaginary growth rate, $z_1 = \tilde z_1 e^{i\omega t}$, $z_2 =  \tilde z_2 e^{i\omega t}$, $w_1 =  \tilde w_1 e^{i\omega t}$, $w_2 =  \tilde w_2 e^{i\omega t}$, we obtain the following equations for $\omega$ and $K$:
\begin{align}
&\left(\frac{K}{2}\right)^2 = -\frac{(1+\omega^2 T^2)^2}{1-\omega^2 T^2}[(\omega-\Omega_1)(\omega-\Omega_2)+\kappa^2],\label{eq:07}\\
&\kappa(2\omega - \Omega_1 -\Omega_2) = \frac{2\omega T[\kappa^2 - (\omega-\Omega_1)(\omega-\Omega_2)]}{1-\omega^2 T^2}.\label{eq:08}
\end{align}
In what follows we will focus on the case $0 < \kappa \ll 1$. (The case of general $\kappa$ is treated exactly for the symmetric case $\Omega_1 = -\Omega_2$ below). To balance Eq.~(\ref{eq:08}) when $\kappa \ll 1$ there are three options, to first order in $\kappa$: $\omega = \Omega_1 + \omega_1 \kappa$, $\omega = \Omega_2 + \omega_2\kappa$, and $\omega = \omega_3 \kappa$,  where $\omega_i$ are constants to be determined. Inserting these in Eqs.~(\ref{eq:07})-(\ref{eq:08}) we find the corresponding values of $K$, to leading order in $\kappa$:
\begin{align}
K_1 &= 2(1+\Omega_1^2T^2)\sqrt{\frac{\kappa(\Omega_1-\Omega_2)}{2\Omega_1 T}},\label{eq:09}\\
K_2 &= 2(1+\Omega_2^2T^2)\sqrt{\frac{\kappa(\Omega_2-\Omega_1)}{2\Omega_2 T}},\label{eq:10}\\
K_3 &= 2\sqrt{-\Omega_1 \Omega_2}.\label{eq:11}
\end{align}
These values of $K$ are real and positive if the frequencies $\Omega_1$ and $\Omega_2$ have opposite sign. In general, depending on the values of $\Omega_1$, $\Omega_2$, and $T$, one can have any ordering of $K_1$, $K_2$, and $K_3$, leading to different bifurcation structures. In Fig.~\ref{fig:02} we obtain the stability diagram for the system by plotting $K_1$ (solid blue), $K_2$ (dashed red), and $K_3$ (dot-dashed black) as a function of $T$ for $\kappa = 0.02$, $\Omega_1 = -1$, and $\Omega_2 = 2$. The sequence of bifurcations shows regions of stability, with local synchronization when $K<K_1,K_2,K_3$, global synchronization when $K>K_3$, asymmetric suppression where $K_1<K<K_2,K_3$ or $K_2<K<K_1,K_3$, and incoherence when $K_1,K_2<K<K_3$. (We note that the region of stability for global synchronization typically stretches below $K_3$, which we will see below.) Regions of stability are labeled in Fig.~\ref{fig:02} and a zoomed-in view of the small-$T$ region is shown in panel (b). We also indicate the time delays $T=1$ and $T=0.1$ used in Figs.~\ref{fig:01}(a) and (b) using vertical dotted lines. Lastly, the interplay between $K_1$ , $K_2$, and $K_3$ illuminates the asymmetric suppression state as follows. First, asymmetric suppression states are only attainable for time delays $T$ where $\text{min}(K_1,K_2)<K_3$. They are then realized when $K$ surpasses either $K_1$ or $K_2$, but not both. The mode that becomes unstable at $K = K_1$ ($K = K_2$) is localized in community $1$ ($2$). More precisely, at $K = K_{1,2}$ the mode with imaginary growth rate satisfies $r_{2,1} \propto \kappa^{1/2} r_{1,2}$ (see Appendix~\ref{QC}), so that for $\kappa \ll 1$ the synchronized mode is localized in one community. When $K_1<K<K_2$ or $K_2<K<K_1$, we find that the values of $r_1$, $r_2$ saturate at values consistent with the mode localization predicted by the linear stability analysis. In particular, when $K_1<K<K_2$ community $1$ reaches near incoherence, whereas when $K_2<K<K_1$ community $2$ reaches near incoherence. Moreover, communities $1$ and $2$ swap roles in the asymmetric suppression state when $K_1$ and $K_2$ intersect at the critical time delay $T_c=\sqrt{\sqrt{\Omega_2}-\sqrt{-\Omega_1}}/\sqrt{\sqrt{-\Omega_1}\Omega_2^2-\sqrt{\Omega_2}\Omega_1^2}$, at which point no asymmetric suppression state exists and the local synchronization state gives way directly to incoherence. Assuming that $|\Omega_2|>|\Omega_1|$ this shows that for sufficiently small $T$ it is community $1$ that desynchronizes first, while for larger $T$ it is community $2$ that desynchronizes first.

\begin{figure}[t]
\centering
\epsfig{file =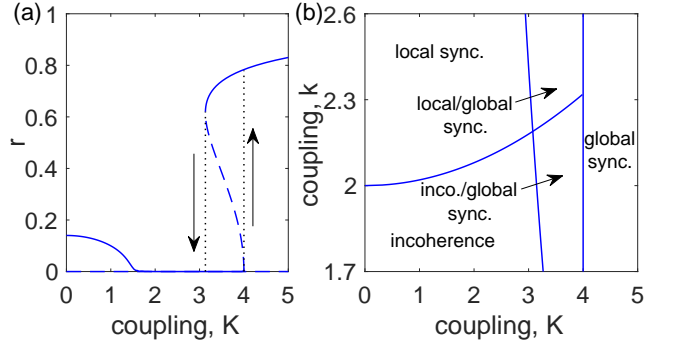, clip =,width=1\linewidth }
\caption{{\it Symmetric case: Two communities.} (a) Local order parameters $r=r_1=r_2$ versus $K$ for $\kappa=0.02$, $\Omega=\Omega_1=-\Omega_2=2$, and $T=1$. Solid and dashed curves indicate stable and unstable solutions; vertical dotted lines with arrows indicate hysteresis. (b) Stability diagram in $(K,k)$ space for the symmetric case, using $\Omega = 2$ and $T=1$.} \label{fig:03}
\end{figure}

To gain further insight into the hysteretic nature of the transition to global synchronization we now focus on the symmetric case, $\Omega_2 = -\Omega_1 = \Omega$. Searching for phase-locked solutions of Eqs.~(\ref{eq:04})--(\ref{eq:06}) of the form $z_1 = r e^{i \psi}$, $z_2 = r e^{i \psi + \alpha}$, with $r$, $\psi$, $\alpha$ constants and $w_{1,2}=z_{1,2}$, gives the following implicit expression for the synchronized branch:
\begin{align}
\left(\frac{K}{2}\right)^2 = \frac{\Omega^2}{(1+r^2)^2} +\left( \frac{r^2}{(1-r^2)}  - \kappa \right)^2. \label{eq:12}
\end{align}
By taking the limit $r\to0^+$ we see that this branch begins at $K=K_3$, as expected. Whether this bifurcation is subcritical or supercritical depends on whether  $K''(0) =  -4\frac{ \kappa + \Omega^2}{ \sqrt{\kappa^2+\Omega^2}}$ is negative or positive, respectively, indicating that bistability exists when $\kappa>-\Omega^2$. Note that an isolated community (i.e., at $K = 0$) is incoherent (synchronized) for  $\kappa < 0$ ($\kappa>0$), and thus the transition at $K_3$ is subcritical if (but not only if) there is a synchronized branch at $K = 0$. In Fig.~\ref{fig:03}(a) we plot the value of $r$ obtained from Eq.~(\ref{eq:12}) together with the locally synchronized and incoherent states, with solid and dashed curves indicating stable and unstable branches. We note that both stable and unstable branches describe globally synchronized states, i.e., states where the communities are phase-locked, but with the unstable branch displaying smaller degree of local synchronization. Conditions for the stability of the incoherent state can be determined using the Routh-Hurwitz criterion on the Jacobian associated with the linearization of Eqs.~(\ref{eq:04})--(\ref{eq:06}) (see Appendix~\ref{two symmetric}). When $0 \leq \kappa \ll 1$, these conditions reduce to
\begin{align}
2(1+\Omega^2T^2)\sqrt{\frac{\kappa}{T}}<K<2 \Omega.\label{eq:13}
\end{align}
In Fig.~\ref{fig:03}(b) we plot these bifurcation curves in the stability diagram in $(K,k)$ space for the symmetric case, using $\Omega = 2$ and $T=1$. Along with the bifurcations indicated by Eq.~(\ref{eq:13}) we also indicate the lower bound (in $K$) for the globally synchronized state obtained from Eq.~(\ref{eq:12}). In addition to incoherence, local synchronization, and global synchronization, we identify bistable regions between incoherence and global synchronization and between local and global synchronization where the globally synchronized branch folds over either of the other states.

\begin{figure}[t]
\centering
\epsfig{file =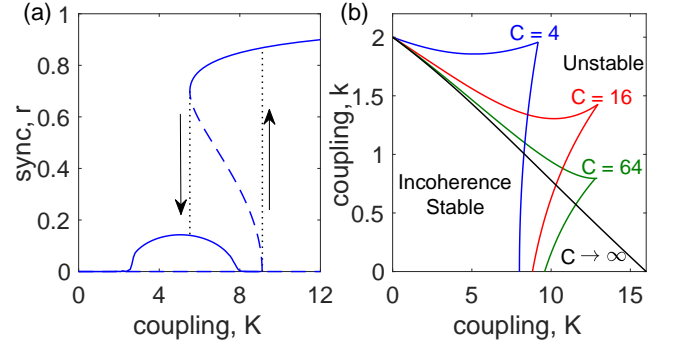, clip =,width=1\linewidth }
\caption{{\it Many communities.} (a) Local order parameters $r=r_\sigma$ versus $K$ for the $C=4$ community case for $\kappa=-0.05$, $T=1$, and $\Omega=\Omega_{\text{even}}=-\Omega_{\text{odd}}=2$. Solid and dashed curves indicate stable and unstable solutions; vertical dotted lines with arrows indicate hysteresis. (b) Stability diagram for the incoherent state in $(K,k)$ space for the $C=4$, $16$, and $64$ cases along with the $C\to\infty$ limit.} \label{fig:04}
\end{figure}

\subsection{Many communities} 

Lastly, we consider larger systems comprised of many ($C>2$) communities that display even richer dynamics. For simplicity we let $C$ be even and consider the case where $\Omega=\Omega_{\text{even}}=-\Omega_{\text{odd}}$. (We note that $C>2$ yields a system that is qualitatively distinct from the $C=2$ case because of the inclusion of time-delayed interactions between even communities and time-delayed interactions between odd communities.)
Even for the $C=4$ case the resulting dynamics are more complicated than the $C=2$ case (presented above), most notably due to a richer sequence of non-monotonic synchronization transitions. We illustrate this in Fig.~\ref{fig:04}(a), where we plot the local order parameters $r=r_\sigma$ versus coupling $K$ for $\kappa = -0.05$, $T=1$, and $\Omega=2$. Since $\kappa<0$ the systems begins in the incoherent state when $K=0$, but as $K$ is increased the system undergoes a first bifurcation to local synchronization followed by a second bifurcation back to incoherence. This is then followed by a third subcritical bifurcation to global synchronization. Decreasing $K$ highlights another hysteresis loop, however the transitions are again more complicated, first with a bifurcation from global synchronization to local synchronization (via explosive desynchronization) then a second bifurcation back to the incoherent state.

A linear stability analysis of the incoherent state (See Appendix~\ref{many communities}) yields the critical values
\begin{align}
K^*&=\frac{2T^2\Omega^2+2(T\kappa-1)^2}{nT(T\kappa-1)}\bigg[(n-1)(T\kappa-1)\nonumber\\
&\pm\sqrt{n^2(T\kappa+1)^2-2n(T\kappa-1)^2+(T\kappa-1)^2}\bigg],\label{eq:14}\\
K^\dagger&=\frac{4\kappa n(n-1)+4n\sqrt{n^2\kappa^2+\Omega^2(2n-1)}}{2n-1},\label{eq:15}
\end{align}
where $n=C/2$. We note that the $\pm$ choice in Eq.~(\ref{eq:14}) corresponds to two different branches of the same curve. The combination of these critical values describes the bifurcation involving the incoherent state, and allows us to sketch the stability diagram for the incoherent state in Fig.~\ref{fig:04}(b) illustrating the $C=4$, $16$, and $64$ cases along with the limiting case $C\to\infty$. The lower-left regions of the diagram represent the respective regions of stability for the incoherent state and highlight the potential for non-monotonic transitions as $K$ is increased for fixed $k$. In particular, for the $C=4$ and $16$ cases the upper portion of the bifurcation curve (given by $K^*$) decreases, then increases, giving rise to non-monotonic transitions similar to that observed in Fig.~\ref{fig:04}(a). This phenomenon does not persist for arbitrarily large numbers of communities, however, as can be seen for the $C=64$ case, where the $K^\dagger$ branch intersects the $K^*$ branch before the minimum is reached. In the $C\to\infty$ limit the bifurcation comes solely from the $K^*$ branch (using the $+$ sign), yielding $K_\infty^*=4\kappa(T^2\Omega^2+(T\kappa-1)^2)/(T\kappa-1)$. Moreover, the symmetry $\Omega_{\text{even}}=\Omega_{\text{odd}}$ allows us to calculate the globally synchronized branch analytically using similar techniques as those used in the two community case (see Appendix~\ref{many communities}), resulting in the implicit equation (defining $K$ in terms of $r$)
\begin{align}
K &= \frac{2}{(2n-1)(1-r^2)}\bigg[(n-1)(\kappa-r^2(1+\kappa))\nonumber\\
&+n^2(1-r^2)\sqrt{\frac{(\kappa-r^2(1+\kappa))^2}{n^2(1-r^2)^2}+\frac{(2n-1)\Omega^2}{n^4(1+r^2)^2}}\bigg].\label{eq:16}
\end{align}
The globally synchronized branch plotted in Fig.~\ref{fig:04}(a) is given by Eq.~(\ref{eq:16}), with solid and dashed curves indicating stable and unstable branches.

\section{Discussion}

Despite the possible presence of both time delays and community structure in a number of real-world systems with synchronization properties, e.g., bacteria~\cite{Prindle2012Nature,Scott2016ACS}, power grids~\cite{Nishikawa2015NJP,Kim2015NJP} and brain dynamics~\cite{Deco2011Frontiers,Sporns2016ARP}, the collective dynamics that emerge from their combination in heterogeneous systems has to date remained relatively unexplored. In this work we have demonstrated that the combination of these two important properties in coupled oscillator systems gives rise to a rich landscape of dynamical phenomena that does not arise from either of these property in isolation. Using both numerical simulations and analytical techniques we have shown that such systems often go through non-monotonic sequences of synchronization transitions, whereby increasing the coupling strength can first inhibit synchronization, and then promote it. In the two community case this manifests in a first bifurcation from local synchronization to incoherence, then a second (subcritical) bifurcation from incoherence to global synchronization. In the presence of more than two communities we demonstrated that this sequence is more complicated, with an initial bifurcation from incoherence to local synchronization, followed by the sequence described above. Moreover, when community-wise parameters are chosen to be asymmetric, the communities asymmetrically suppress one another's synchronization properties, leading one community to win as it forces the other to near incoherence. Interestingly, the roles of the two communities, in terms of which one remains synchronized while the other is pushed to near incoherence, unexpectedly reverse depending on the time delay. The novel phenomena observed in this work demonstrates the rich dynamics that can emerge from different dynamical and structural properties in oscillator systems.

\appendix


\section{Derivation of the Low Dimensional Equations}\label{derivation}

First we present the derivation of the low dimension equations, namely Eqs.~(2) and (3), from the full high dimensional system given by Eq.~(1). We begin by considering the continuum limit $N_\sigma\to\infty$ taken in such a way that the fraction of oscillator in each community remains constant. In this limit the macroscopic system state of each community $\sigma$ can be described by the density function $f_\sigma(\theta,\omega,t)$ that describe the fraction of oscillators in each community $\sigma$ with phase between $\theta$ and $\theta+d\theta$ and frequency between $\omega$ and $\omega+d\omega$ at time $t$. We seek to describe the macroscopic dynamics as described by the local order parameters $z_\sigma=N_\sigma^{-1}\sum_{j=1}^{N_{\sigma}}e^{i\theta_j^{\sigma}}$. We also introduce the set of time delayed order parameters $w_i^{\sigma\sigma'}=N_{\sigma'}^{-1}\sum_{j=1}^{N_{\sigma'}}e^{i\theta^{\sigma'}(t-\tau_{ij}^{\sigma\sigma'})}$. In the continuum limit the order parameters can be written
\begin{align}
z_\sigma = \iint f_\sigma(\theta,\omega^\sigma,t)d\theta d\omega^\sigma,\label{eq:a01}
\end{align}
and
\begin{align}
w_i^{\sigma\sigma'} = \int z_{\sigma'}(t-\tau) h_{\sigma{\sigma'}}(\tau)d\tau \equiv w_{\sigma'}\label{eq:a02},
\end{align}
where we used the fact that the time delays $\tau_{ij}^{\sigma\sigma'}$ in the definition of $w_i^{\sigma\sigma'}$ are chosen from the distribution  $h_{\sigma\sigma'}(\tau)\propto s^{-\tau/T_{\sigma'}}$, which is independent of $i$ and $\sigma$.

Next, due to the conservation of oscillators, the density functions each satisfy their own continuity equation
\begin{widetext}
\begin{align}
0&=\frac{\partial}{\partial t}f_\sigma+\frac{\partial}{\partial \theta}\left(f_\sigma\dot{\theta}^\sigma\right)\nonumber\\
&=\frac{\partial}{\partial t}f_\sigma+\frac{\partial}{\partial \theta}\left[f_\sigma\left(\omega^\sigma+\frac{K^{\sigma\sigma}}{2i}\left(z_\sigma e^{-i\theta^\sigma}-z_\sigma^*e^{i\theta^\sigma}\right)+\sum_{\sigma'\ne\sigma}^C\frac{K^{\sigma\sigma'}}{2i}\left(w_{\sigma'}e^{-i\theta^\sigma}-w_{\sigma'}^*e^{i\theta^\sigma}\right)\right)\right].\label{eq:a03}
\end{align}
\end{widetext}
Since the natural frequencies remain fixed and are drawn from the distributions $g_{\sigma}(\omega^\sigma)$, the density functions can be written in a Fourier series of the form
\begin{align}
f_\sigma(\theta,\omega^\sigma,t)=\frac{g_\sigma(\omega^\sigma)}{2\pi}\left[1+\sum_{n=1}^\infty \hat{f}_{\sigma,n}(\omega^\sigma,t)e^{in\theta}+\text{c.c.}\right],\label{eq:a04}
\end{align}
where c.c. stands for the complex conjugate of the previous term. Following the dimensionality reduction technique of Ott and Antonsen~\cite{Ott2008Chaos,Ott2009Chaos} we propose the ansatz that the Fourier coefficients decay geometrically, i.e.,
\begin{align}
f_\sigma(\theta,\omega^\sigma,t)=\frac{g_\sigma(\omega^\sigma)}{2\pi}\left[1+\sum_{n=1}^\infty a_\sigma^n(\omega^\sigma,t)e^{in\theta}+\text{c.c.}\right].\label{eq:a05}
\end{align}
Remarkably, the dynamics defined by the infinite collection of Fourier coefficients collapse onto the low dimensional manifold describing the evolution of $a$ via the differential equation
\begin{align}
\partial_t a_\sigma = -i\omega^\sigma a_\sigma +\frac{k}{2}\left(z_\sigma^*-z_\sigma a_\sigma^2\right)+\frac{K}{C}\sum_{\sigma'\ne\sigma}\left(w_{\sigma'}^*-w_{\sigma'} a_\sigma^2\right),\label{eq:a06}
\end{align}
where we have used $K^{\sigma\sigma}=k$ and $K^{\sigma\sigma'}=2K/C$ (for $\sigma\ne\sigma'$).

Next, the function $a_\sigma$ can be tied directly to the local order parameter by inserting Eq.~(\ref{eq:a05}) into Eq.~(\ref{eq:a01}), yielding
\begin{align}
z_\sigma=\int g_\sigma(\omega^\sigma)a_\sigma^*(\omega^\sigma,t)d\omega^\sigma.\label{eq:a07}
\end{align}
Assuming that the frequency distributions are Lorentzian, i.e., $g_\sigma(\omega^\sigma)=\Delta/\{\pi[\Delta^2+(\omega^\sigma-\Omega_\sigma)^2]\}$, the integral can be evaluated using Cauchy's residue theorem, yielding $z_\sigma = a_\sigma^*(\Omega_\sigma-i\Delta,t)$. Evaluating Eq.~(\ref{eq:a06}) at $\omega^\sigma=\Omega_\sigma-i\Delta$ and taking a complex conjugate yields
\begin{align}
\dot{z}_\sigma&=(-\Delta+i\Omega_\sigma)z_\sigma + \frac{k}{2}(z_\sigma-z_\sigma^*z_\sigma^2) + \frac{K}{C}\sum_{\sigma'\ne\sigma}(w_{\sigma'}-w_{\sigma'}^*z_\sigma^2).\label{eq:a08}
\end{align}

Lastly, to close the dynamics we convert Eq.~(\ref{eq:a02}) to a differential equation for $w_\sigma$. Taking a Fourier transform, we have that
\begin{align}
(1+T_{\sigma}s)\overline{w}_\sigma(s)=\overline{z}_\sigma(s), \label{eq:a09}
\end{align}
and converting back to the time domain yields
\begin{align}
T_\sigma \dot{w}_\sigma&=z_\sigma-w_\sigma.\label{eq:a10}
\end{align}
thereby closing the dynamics.

\begin{figure}[ht]
\centering
\epsfig{file =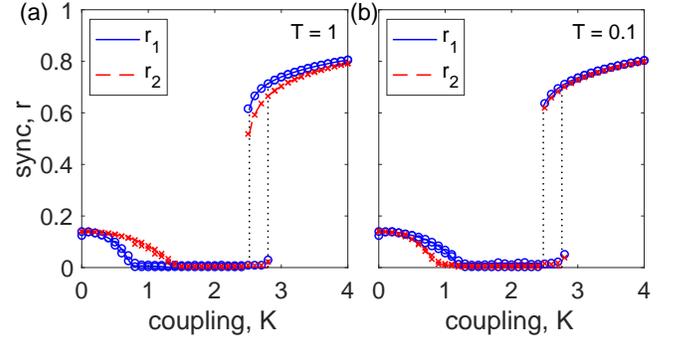, clip =,width=1\linewidth }
\caption{{\it Numerical validation: Asymmetric case.} Local order parameters $r_1$ (solid blue) and $r_2$ (dashed red) versus $K$ for $\kappa=0.02$, $\Omega_1=-1$, $\Omega_2=2$, and time delays $T=1$ (a) and $0.1$ (b). Results from direct simulations with $N=2\times10^4$ oscillators in each community are plotted in blue circles and red crosses.}\label{fig:a01}
\end{figure}

\section{Numerical Validation of the Low Dimensional Equations}\label{validation}

Next we present simulations validating the low dimensional equations. Since it is computationally infeasible for us to account for both a distribution of time delays and community structure, we use the hybrid approach proposed in Ref.~\cite{Lee2011Chaos}. In this approach, the microscopic dynamics of the oscillators is taken into account, but the effect of the time delay is accounted via the techniques presented above. We note that the treatment of distributed time delays via the Ott-Antonsen Ansatz has been validated before  \cite{Lee2009PRL}, so we focus on validating its interplay with the community structure. The hybrid approach consists in considering the equations
\begin{align}
\dot{\theta}_i^\sigma &= \omega_i^\sigma + kr_\sigma\sin(\psi_\sigma-\theta_i^\sigma) + K\sum_{\sigma'\ne\sigma}\rho_{\sigma'}\sin(\phi_{\sigma'}-\theta_i^\sigma),\label{eq:a11}\\
\dot{w}_\sigma &= (z_\sigma-w_\sigma)/T_\sigma,\label{eq:a12}
\end{align}
where $z_\sigma=r_\sigma e^{i\psi_\sigma}$, $w_\sigma=\rho_\sigma e^{i\phi_\sigma}$, and we have taken advantage of the technique used above to deal with the time delays. Note that the time delays enter in the evolution of oscillator $i$ by the fact that the input it receives from community $\sigma'$, $w_{\sigma'}$, is effectively a delayed version of the order parameter from community $\sigma'$. Both Eqs.~(\ref{eq:a11})-(\ref{eq:a12}) and Eqs.~(1) have the same low-dimensional description, Eqs.~(\ref{eq:a08})-(\ref{eq:a10}). For a further discussion of the hybrid approach, see Ref.~\cite{Lee2011Chaos}.

We begin by considering the asymmetric two community case with mean frequencies $\Omega_1=-1$ and $\Omega_2=2$ and $\kappa=0.02$. In Fig.~\ref{fig:a01} we plot the steady-state values of $r_1$ and $r_2$ in solid blue and dashed red curves, obtained in the same way to produce Fig.~1, with time delays $T=1$ and $0.1$ in panels (a) and (b). We then overlay the results from direct simulations of Eqs.~(\ref{eq:a11})-(\ref{eq:a12}) with $N=2\times10^4$ oscillators in each community in blue circles and red crosses. We note that the results from simulations are in excellent agreement.

\begin{figure}[ht]
\centering
\epsfig{file =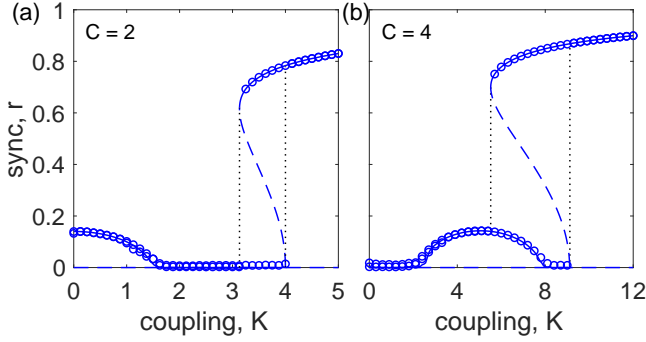, clip =,width=1\linewidth }
\caption{{\it Numerical validation: Symmetric case.} Local order parameters $r=r_\sigma$ versus $K$ with $T=1$, for (a) the two community case with $\kappa=0.02$ and $(\Omega_1,\Omega_2)=(-2,2)$, and (b) the four community case with $\kappa=-0.05$ and $(\Omega_1,\Omega_2,\Omega_3,\Omega_4)=(-2,2,-2,2)$. Results from direct simulations with $N=2\times10^4$ oscillators in each community are plotted in blue circles.}\label{fig:a02}
\end{figure}

Next we consider the symmetric case with two and four communities with $(\Omega_1,\Omega_2)=(-2,2)$ and $(\Omega_1,\Omega_2,\Omega_3,\Omega_4)=(-2,2,-2,2)$, respectively, and $T=1$. In Fig.~\ref{fig:a02}(a) and (b) we plot the steady-state values of $r_1$ (which are indistinguishable from $r_2$, $r_3$, and $r_4$) in blue for the two and four community cases, respectively. Solid and dashed curves indicating stable and unstable branches, obtained in the same way to produce Fig.~3(a) and Fig.~4(a). We then overlay the results from direct simulations with $N=2\times10^4$ oscillators in each community in blue circles, again noting excellent agreement.

\section{Linear Stability Analysis of the Incoherent State}\label{linear stability analysis}

Here we present linear stability analyses of the incoherent state for different mean frequency configurations. We restrict our attention to an even number of communities, $\sigma = 1, 2 \dots , 2n.$ 

\subsection{Two communities with symmetric frequencies: exact analysis}\label{two symmetric}
First, we consider the case of two communities, $n = 1$, with symmetric mean frequencies, i.e., $\Omega_1 = -\Omega_2 \equiv \Omega$. The Jacobian associated with perturbations from the incoherent solution $z_1 = z_2 = w_1 = w_2 = 0$ of Eqs.~(\ref{eq:04})-(\ref{eq:06}) is
\begin{align}
J = \left(
\begin{array}{cccc}
 \kappa +i \Omega  & 0 & 0 & \frac{K}{2} \\
 0 & \kappa -i \Omega  & \frac{K}{2} & 0 \\
 \frac{1}{T} & 0 & -\frac{1}{T} & 0 \\
 0 & \frac{1}{T} & 0 & -\frac{1}{T} \\
\end{array}
\right)\nonumber
\end{align}
The characteristic equation is given by $a_0 + a_1 \lambda + a_2 \lambda^2 +a_3 \lambda^3 +a_4 \lambda^4 =  0$, with 
\begin{align}
a_0 &= 4 \kappa ^2-K^2+4 \Omega ^2, \nonumber\\
a_1 &= 8   \left(-\kappa +\kappa ^2 T+T \Omega ^2\right),\nonumber\\
a_2 &= 4\left(\kappa ^2 T^2+T^2 \Omega ^2-4 \kappa  T+1\right),\nonumber\\
a_3 &= 8  T (1-\kappa  T),\nonumber\\
a_4 &= 4 T^2.\nonumber
\end{align}
Applying the Routh-Hurwitz criterion \cite{Hurwitz1895}, we obtain necessary and sufficient conditions for linear stability. The eigenvalues have all negative real part, corresponding to linear stability of the incoherent state, if the Routh-Hurwitz inequalities $a_i > 0$, $a_3 a_2 > a_4 a_1$, and $a_1 a_2 a_3 > a_1^2 a_4 + a_0 a_3^2$ are satisfied. These inequalities simplify to
\begin{align}
&T > 0,\nonumber\\
&\kappa T < 1,\nonumber\\
& T(\Omega^2 + \kappa^2) > \kappa,\nonumber\\
&K < 2 \sqrt{\Omega^2 + \kappa^2},\nonumber\\
&1 + T^2(\Omega^2 + \kappa^2) > 4 \kappa T,\nonumber\\
& \kappa T [4 + T^2(\Omega^2 + \kappa^2)] < 1+4\kappa^2 T^2,\nonumber\\
&K >  \frac{2 \sqrt{\kappa} \left[(1-\kappa T)^2+T^2 \Omega^2\right]}{\sqrt{T}(1 - \kappa T)}.\nonumber
\end{align}
When $T \to 0$ with $\kappa > 0$ the inequalities are not satisfied, showing that time delay is necessary to stabilize the incoherent state. For a given positive time delay, and sufficiently small $\kappa$, the inequalities reduce to $K_{1,2} < K < K_3$, with 
\begin{align}
&K_{1,2} = \frac{2 \sqrt{\kappa} \left[(1-\kappa T)^2+T^2 \Omega^2\right]}{\sqrt{T}(1 - \kappa T)},\nonumber\\
&K_3 = 2 \sqrt{\Omega^2 + \kappa^2}.\nonumber
\end{align}
To leading order in $\kappa$, these inequalities reduce to Eq.~(13),
\begin{align}
2(1+\Omega^2T^2)\sqrt{\frac{\kappa}{T}}<K<2 \Omega, \nonumber
\end{align}
and agree with Eqs.~(8)-(10) when $\Omega_1 =  -\Omega_2$.

\subsection{Even communities with symmetric frequencies}\label{many communities}

Here we consider the case of an even number $C = 2n$ of communities with mean frequencies $\Omega_{\sigma} = \Omega$, $\sigma= 1,2,\dots,n$, $\Omega_{\sigma} = -\Omega$, $\sigma= n+1,\dots,2n$, which corresponds to the case shown in Fig.~(4). Linearizing Eqs.~(\ref{eq:08})-(\ref{eq:10}) about the incoherent state we obtain
\begin{align}
&\dot{\delta z}_\sigma = \kappa \delta z_\sigma + i\Omega \delta z_\sigma+ \frac{K}{2n} \sum_{\sigma' \neq \sigma}\delta w_{\sigma'},\hspace{0.5cm} \sigma = 1, \dots, n,\label{eq:zplus}\\
&\dot{\delta z}_\sigma = \kappa \delta z_\sigma - i\Omega \delta z_\sigma+ \frac{K}{2n} \sum_{\sigma' \neq \sigma}\delta w_{\sigma'},\hspace{0.5cm} \sigma = n+1, \dots, 2n,\label{eq:zminus}\\
&T \dot{\delta w}_\sigma = \delta z_\sigma - \delta w_\sigma,\hspace{3.3cm} \sigma = 1, \dots, 2n.\label{eq:wplusminus}
\end{align}
Anticipating that communities with the same frequency will remain synchronized, we introduce the variables 
\begin{align}
z_+ &= \sum_{\sigma=1}^n \delta z_\sigma, \hspace{1cm} w_+ = \sum_{\sigma=1}^n \delta w_\sigma, \\
z_- &= \sum_{\sigma=n+1}^{2n} \delta z_\sigma, \hspace{1cm} w_- = \sum_{\sigma=n+1}^{2n} \delta w_\sigma,\nonumber\\
x_j &=\delta z_1 - \delta z_j, \hspace{0.2cm} y_j = \delta w_1 - \delta w_j, \hspace{0.8cm} j = 2,\dots,n\nonumber\\
x_j &=\delta z_{n+1} - \delta z_j, \hspace{0.1cm} y_j = \delta w_{n+1} - \delta w_j, \hspace{0.2cm} j = n+2,\dots,2n.\nonumber
\end{align}
Perturbations $z_+$ and $z_-$ are our main interest. We will show that the perturbations $x_j$, $y_j$ decay, implying that the only relevant synchronization modes are those in which communities with equal frequencies behave identically. 
Inserting the new variables in Eqs.~(\ref{eq:zplus})-(\ref{eq:wplusminus}) we obtain the decoupled systems
\begin{align}
&\dot z_+ = \kappa z_+ + i\Omega z_+ + \frac{K}{2n}[(n-1) w_+  + n  w_-], \label{trans1}\\
&T \dot w_+ = z_+ - w_+,\\
&\dot z_- = \kappa z_- - i\Omega z_- + \frac{K}{2n}[(n-1) w_-  + n  w_+],\\ 
&T \dot w_- = z_- - w_-, \label{trans2}
\end{align} 
and
\begin{align}
&\dot x_j = \kappa x_j + i\Omega x_j - \frac{K}{2n} y_j,\label{xy1}\\ 
&T \dot y_j = x_j - y_j, \hspace{2cm} j = 2,\dots,n,\\
&\dot x_j = \kappa x_j - i\Omega x_j  - \frac{K}{2n} y_j,\\
&T \dot y_j = x_j - y_j, \hspace{2cm} j = n+2,\dots,2n. \label{xy2}
\end{align} 
First we focus on the perturbations that leave communities with the same mean frequencies synchronized, i.e., $z_{\pm}$ and $w_{\pm}$. We determine the location of potential bifurcations by finding the values of $K$ where the growth rate associated to Eqs.~(\ref{trans1})-(\ref{trans2}) is purely imaginary. We set $z_{\pm}(t) = e^{i \omega t}\tilde z_{\pm}$, $w_{\pm}(t) = e^{i \omega t} \tilde w_{\pm}$ and obtain
\begin{align}
&i \omega \tilde z_+ = \kappa \tilde z_+ + i\Omega \tilde z_++ \frac{K}{2n}((n-1)\tilde w_+ +n \tilde w_-), \nonumber\\
 &i \omega T  \tilde w_+ = \tilde z_+ - \tilde w_+,\nonumber\\
&i \omega \tilde z_- = \kappa \tilde z_- - i\Omega \tilde z_-  + \frac{K}{2n}((n-1)\tilde w_- +n \tilde w_+),\nonumber\\
 &i \omega T  \tilde w_- = \tilde z_- - \tilde w_-.\nonumber
\end{align} 
Eliminating $\tilde z_{\pm}$ and $\tilde w_{\pm}$ leads to the complex equation for $K$ and $\omega$
\begin{widetext}
\begin{align}
\frac{K^2}{4(1+i\omega T)^2} = \left[i(\omega - \Omega)- \kappa - \frac{K (n-1)}{2n(1+ i\omega T)}\right] \left[i(\omega + \Omega)- \kappa - \frac{K (n-1)}{2n(1+ i\omega T)}\right].\nonumber
\end{align}
\end{widetext}
Separating the equation into real and imaginary parts and solving the algebraic equations, we find the critical values
\begin{widetext}
\begin{align}
K^*&=\frac{2T^2\Omega^2+2(T\kappa-1)^2}{nT(T\kappa-1)}\bigg[(n-1)(T\kappa-1)\pm\sqrt{n^2(T\kappa+1)^2-2n(T\kappa-1)^2+(T\kappa-1)^2}\bigg],\label{eq:a14}\\
K^\dagger&=\frac{4\kappa n(n-1)+4n\sqrt{n^2\kappa^2+\Omega^2(2n-1)}}{2n-1},\label{eq:15}
\end{align}
\end{widetext}
To complete the analysis, we need to consider the evolution of perturbations transversal to the manifold where communities with the same frequency are synchronized, i.e., $x_{j}$, $y_{j}$. Setting $x_j(t) = e^{i \alpha t}\tilde x_j$, $y_j(t) = e^{i \alpha t} \tilde y_j$ in Eqs.~(\ref{xy1})-(\ref{xy2}) and eliminating $\tilde x_j$, $\tilde y_j$, we obtain the critical values
\begin{align}
K_T &= 4\kappa \left[ 1 + \left(\frac{\Omega T }{1 - \kappa T}\right)^2 \right],\nonumber\\
\omega &= \frac{\Omega}{1-\kappa T}.\nonumber
\end{align}
From the exact form of the eigenvalues associated to Eqs.~(\ref{xy1})-(\ref{xy2}), one can check that their real part becomes negative for $K> K_T$. For $\kappa > 0$, the incoherent state is already incoherent [the cusp in the curves in Fig.~4(b) occurs at $\kappa = (\sqrt{2n-1}-n)/(n+\sqrt{2n-1}) < 0$]. For the region of interest, $\kappa < 0$, we obtain then $K_T < 0$, which means that the only relevant modes of desynchronization are those in which communities with the same mean frequency remain synchronized, which yield the critical values in Eqs.~(\ref{eq:a14}),~(\ref{eq:15}) plotted in Fig.~4(b).

In addition to the linear stability of the incoherent state, we present here the derivation of Equation~(16) for the globally synchronized branch. We look for steady state solutions of (\ref{eq:a08})-(\ref{eq:a10}) such that communities with the same frequency have the same order parameters, and set $z_\sigma = w_\sigma = r$ for $\sigma = 1,2\dots,n$, $z_\sigma = w_\sigma = r e^{i \alpha}$ for $\sigma = n+1,\dots,2n$, where $r$ and $\alpha$ are constants. This Ansatz gives
\begin{align}
&0 = i\Omega r + \kappa r -(1+\kappa)r^3 +\frac{K}{2n}(n-1)(r-r^3)  \nonumber\\
&+\frac{K}{2}(r e^{i\alpha} - r^3 e^{-i\alpha}).
\end{align}
Separating the equation into real and imaginary parts and solving for $K$ gives Eq.~(16) in the main text,
\begin{align}
K &= \frac{2}{(2n-1)(1-r^2)}\bigg[(n-1)(\kappa-r^2(1+\kappa))\nonumber\\
&+n^2(1-r^2)\sqrt{\frac{(\kappa-r^2(1+\kappa))^2}{n^2(1-r^2)^2}+\frac{(2n-1)\Omega^2}{n^4(1+r^2)^2}}\bigg].
\end{align}

\section{Two communities with asymmetric frequencies: Asymmetric suppression}\label{QC}

In this section we consider the case of two communities with asymmetric frequencies, i.e., $\Omega_2 \neq -\Omega_1$. A linear stability analysis of the incoherent state allows us to find the critical coupling constants and the modes of instability that give rise to the asymmetrically suppressed states. 

\subsection{Critical coupling constants}

Linearizing Eqs.~(\ref{eq:a08})-(\ref{eq:a10}) about the incoherent state, and again defining $k/2 = 1+ \kappa$ and setting $\Delta = 1$ we obtain
\begin{align}
&\dot{\delta z_1} = \kappa \delta z_1 + i\Omega_1 \delta z_1 +\frac{K}{2}\delta w_2,\nonumber \\
&\dot{\delta z_2} = \kappa \delta z_2 + i\Omega_2 \delta z_2 +\frac{K}{2}\delta w_1,\nonumber \\
&T \dot{\delta w_1} = \delta z_1 - \delta w_1,\nonumber\\
&T \dot{\delta w_2} =\delta z_2 - \delta w_2.\nonumber
\end{align}
We find the location of bifurcations by looking for purely imaginary growth rates. To this end, we set $\delta z_{1,2} = e^{i \omega t}\tilde z_{1,2}$, $\delta w_{1,2} = e^{i \omega t}\tilde w_{1,2}$, and obtain
\begin{align}
&i \omega \tilde z_1 = \kappa \tilde z_1 + i\Omega_1 \tilde z_1 +\frac{K}{2}\tilde w_2,\label{z1} \\
&i \omega \tilde z_2 = \kappa \tilde z_2 + i\Omega_2 \tilde z_2 +\frac{K}{2}\tilde w_1,\label{z2} \\
&i \omega T \tilde w_1 = \tilde z_1 - \tilde w_1,\label{w1}\\
&i \omega T \tilde w_2 =\tilde z_2 - \tilde w_2. \label{w2}
\end{align}
Eliminating $\tilde z_{1,2}$ and $\tilde w_{1,2}$ we obtain a complex equation for $K$ and $\omega$, with its real and imaginary parts giving
\begin{align}
\left(\frac{K}{2}\right)^2\frac{1-\omega^2T^2}{(1+\omega^2 T^2)^2} &= -(\omega - \Omega_1)(\omega - \Omega_2) + \kappa^2,\label{eq:K1}\\
\left(\frac{K}{2}\right)^2\frac{2 \omega T}{(1+\omega^2 T^2)^2} & = \kappa(2\omega -\Omega_1 - \Omega_2).
\end{align}
Eliminating $K$ from these equations we obtain a cubic equation for $\omega$,
\begin{align}
\kappa(2\omega -\Omega_1 - \Omega_2)(1-\omega^2 T^2) = 2\omega T [\kappa^2 - (\omega - \Omega_1)(\omega - \Omega_2)].\label{eq:cubic}
\end{align}
In what follows we will consider the limit $\kappa \ll 1$ and argue that the three solutions to the cubic are, to leading order, $\omega \approx \Omega_1$, $\omega \approx \Omega_2$, and $\omega \approx \kappa (\Omega_1+\Omega_2)/(2T \Omega_1\Omega_2)$. When $\kappa \ll 1$, the leading order terms in (\ref{eq:cubic}) are 
\begin{align}
\kappa(2\omega -\Omega_1 - \Omega_2)(1-\omega^2 T^2) = -2\omega T (\omega - \Omega_1)(\omega - \Omega_2).\label{eq:orderkappa}
\end{align}
Since the left hand side is of order $\kappa$, the right hand side needs to be of order $\kappa$, which can be accomplished by the three choices $\omega = \Omega_1 + \omega_1 \kappa + \mathcal{O}(\kappa^2)$, $\omega = \Omega_2 + \omega_2 \kappa + \mathcal{O}(\kappa^2)$, or $\omega = \omega_3 \kappa + \mathcal{O}(\kappa^2)$, where $\omega_i$ are to be determined.

First consider the case $\omega = \Omega_{1,2} + \omega_{1,2} \kappa + \mathcal{O}(\kappa^2)$. Inserting this in (\ref{eq:orderkappa}) we obtain
\begin{align}
\omega_{1,2} = \frac{\Omega_{1,2}^2 T^2-1}{2\Omega_{1,2} T},
\end{align}
which, when inserted in Eq.~(\ref{eq:K1}) gives, to leading order in $\kappa$,
\begin{align}
K_1 &= 2(1+\Omega_1^2T^2)\sqrt{\frac{\kappa(\Omega_1-\Omega_2)}{2\Omega_1 T}},\label{eq:KK1}\\
K_2 &= 2(1+\Omega_2^2T^2)\sqrt{\frac{\kappa(\Omega_2-\Omega_1)}{2\Omega_2 T}}.\label{eq:KK2}
\end{align}
Now consider the case $\omega = \omega_3 \kappa + \mathcal{O}(\kappa^2)$. We obtain
\begin{align}
\omega_{3} =(\Omega_1+\Omega_2)/(2T \Omega_1\Omega_2),
\end{align}
and inserting $\omega = \omega_3 \kappa$ in Eq.~(\ref{eq:K1}) gives, to leading order,
\begin{align}
K_3 &= 2\sqrt{-\Omega_1 \Omega_2}.\label{eq:KK0}
\end{align}
For $K = 0$, the communities are decoupled, and since there is no intra-community delay, each community synchronizes at $k = 2$ (corresponding to $\kappa = 0$). Therefore the incoherent state is unstable for small enough positive  $\kappa$, and loses stability when $K$ first increases beyond the smallest of $K_1$, $K_2$, and $K_3$.

\subsection{Asymmetric suppression}

As discussed in the main text, for small $\kappa$ we find that when $K_1 < K < K_2 < K_3$ community $2$ synchronizes while community $1$ remains almost completely incoherent. (An analogous situation occurs if $K_2 < K < K_1 < K_3$.) This can be understood from the linear stability analysis by determining the form of the modes of instability at the bifurcations that occur at $K = K_1$ and $K = K_2$. From Eqs.~(\ref{z1})-(\ref{w2}) we obtain
\begin{align}
\frac{\tilde z_2}{\tilde z_1} = \frac{2 \kappa}{K} [i(\omega - \Omega_1) - \kappa](1+i \omega T).
\end{align}
At $K = K_1$, $\omega = \Omega_1 + \omega_1 \kappa$, and we get to leading order in $\kappa$
\begin{align}
\left|\frac{\tilde z_2}{\tilde z_1} \right| =  \frac{\kappa^{1/2} }{|\Omega_1 - \Omega_2|} \left(\frac{1+\Omega_1^2 T^2}{2\Omega_1 T}\right)^{1/2}.
\end{align}
Similarly, at $K = K_2$ we get $|\tilde z_1| \sim \kappa^{1/2} |\tilde z_2|$. For the parameters used in Figure 1 (a) in the main text, we obtain $|\tilde z_1| \approx 0.053 |\tilde z_2|$ at $K = K_2$, and for those used in Figure 1 (b) we obtain $|\tilde z_1| \approx 0.11 |\tilde z_2|$ at $K = K_1$. For the case of Figure 1(a), where  $K_1 < K < K_2 < K_0$, the mode that desynchronizes as $K$ is decreased past $K_2$ is therefore localized mostly in community $2$. We find that this linear analysis still describes the behavior of the two communities in the interval $K_1 < K < K_2$, where community $2$ remains synchronized while community $1$ has a very small degree of synchronization.


\bibliographystyle{apa}

\end{document}